\documentclass[aps, prd, twocolumn, lengthcheck, superscriptaddress, showpacs, letterpaper, nofootinbib]{revtex4-1}
\usepackage{amsfonts}
\usepackage{amsmath}
\usepackage{amssymb}
\usepackage{xcolor}
\usepackage{graphicx}
\usepackage{longtable}
\usepackage{slashed}
\usepackage{hyperref}
\usepackage{amsthm}
\usepackage{mathrsfs}
\usepackage{color}

\def\be{\begin{eqnarray}}\def\ee{\end{eqnarray}}

\allowdisplaybreaks

\begin{document}
\title{Chiral effective Lagrangian for excited  heavy-light mesons from QCD}

\author{Qing-Sen Chen}
\affiliation{Center for Theoretical Physics, College of Physics, Jilin University,
Changchun 130012, China}

\author{Hui-Feng Fu}
\email{huifengfu@jlu.edu.cn}
\affiliation{Center for Theoretical Physics, College of Physics, Jilin University,
Changchun 130012, China}

\author{Yong-Liang Ma}
\email{ylma@ucas.ac.cn}
\affiliation{School of Fundamental Physics and Mathematical Sciences,
Hangzhou Institute for Advanced Study, UCAS, Hangzhou, 310024, China}
\affiliation{International Center for Theoretical Physics Asia-Pacific (ICTP-AP) (Beijing/Hangzhou), UCAS, Beijing 100190, China}

\author{Qing Wang}
\email{wangq@mail.tsinghua.edu.cn}
\affiliation{Department of Physics, Tsinghua University, Beijing 100084, China
}
\affiliation{Center for High Energy Physics, Tsinghua University, Beijing 100084, China
}


\begin{abstract}
We derive the chiral effective Lagrangian for excited heavy-light mesons from QCD under proper approximations. We focus on the chiral partners with $j_l^P=\frac{3}{2}^+$ and $j_l^P=\frac{3}{2}^-$ which amounts to ($1^+,2^+$) and ($1^-,2^-$) states respectively. The low energy constants including the masses of the chiral partners are calculated. The calculated spectrum for the excited mesons are found roughly consistent with experimental data. In addition, our results indicate that quantum numbers of $B_J(5970)$ can be identified with $1^-$ or $2^-$.
\end{abstract}
%
 \maketitle

\section{Introduction}

The heavy quark spin-flavor symmetry that exact under the $m_Q\rightarrow \infty$ limit with $m_Q$ being the heavy quark mass plays an important role in the hadronic systems containing one heavy quark/antiquark~\cite{Isgur:1989ed,Isgur:1989vq,Eichten:1989zv}. Due to this symmetry, the total angular momentum of a heavy-light meson has eigenvalues $j_{\pm} = j_l \pm 1/2$ with $\vec{j}_l$ being the angular momentum of the light degrees of freedom, and the corresponding eigenstates form a degenerate doublet $(j_-,j_+)$ for each $j_l$. The angular momentum $\vec{j}_l$ can be decomposed as $\vec{j}_l = \vec{s} + \vec{l}$, where $\vec{s}$ is the spin of light quark/antiquark and $\vec{l}$ is the orbital angular momentum. For S-wave heavy-light mesons, $l = 0$ and $j_l^P=\frac{1}{2}^-$ represents the degenerate doublet $(j^P_-,j^P_+ ) = (0^-, 1^-)$. For P-wave, $l = 1$ and $j_l^P=\frac{1}{2}^+$ or $j_l^P= \frac{3}{2}^+$ represent two distinct doublets $(j^P_-,j^P_+ ) = (0^+, 1^+)$ or $(j^P_-,j^P_+ ) = (1^+, 2^+)$, respectively. For D-wave , $l = 2$ and $j_l^P=\frac{3}{2}^-$ or $j_l^P= \frac{5}{2}^-$,  we have two distinct doublets $(j^P_-,j^P_+ ) = (1^-, 2^-)$ and $(j^P_-,j^P_+ ) = (2^-, 3^-)$. The similar argument applies to other excited states~\cite{Yuan:1994iv}.

For the light quarks, the chiral symmetry, which is dynamically (and explicitly) broken, is essential. If the chiral symmetry were preserved, the heavy-light mesons with the same $j_l$ but opposite parities would be degenerated. We call them chiral partners. The mass splitting between the chiral partners reflects the chiral symmetry breaking~\cite{Nowak:1992um,Bardeen:1993ae}.

Since heavy-light mesons capture the features of both heavy quark symmetry and chiral symmetry, it is ideal to study heavy-light meson phenomena using the chiral Lagrangian incorporating heavy quark symmetry. Since 2000, a series of researches on deriving the chiral effective Lagrangian from QCD have been carried out in Refs.~\cite{Wang:1999cp,Yang:2002hea,Jiang:2009uf,Jiang:2010wa,Jiang:2012ir,Jiang:2015dba,Wang:2000mu,Wang:2000mg,Ren:2017bhd}. The advantage of this method is that it can establish the analytic relationships between the low energy constants (LECs) of the chiral effective theory and QCD. Recently, we used the same methodology to derive the chiral Lagrangian incorporating heavy quark symmetry from QCD to study heavy-light mesons~\cite{Chen2020jiq,Chen2020epl}. In these works, we focused on the chiral partners $j_l^P=\frac{1}{2}^-$ and $j_l^P=\frac{1}{2}^+$. The low energy constants of the effective Lagrangian are expressed in terms of the light quark self-energy which can be calculated by using Dyson-Schwinger equations or lattice QCD. Numerical results of the low energy constants are globally in agreements with the experiment data.

In recent years, more and more excited heavy-light mesons have been observed. In the charm sector, many new excited states such as $D_0(2550)$, $D^\ast_J(2680)$, $D(2740)$, $D^{*}_{2}(3000)$, $D_{J}(3000)$, $D_{J}^{*}(3000)$, etc., have been announced by LHCb
\cite{Aaij:2013sza,Aaij:2016fma,Aaij:2019sqk} and BABAR \cite{delAmoSanchez:2010vq}. In the bottom sector, new bottom states $B_{J}(5721)$, $B_{2}^{*}(5747)$, $B_{J}(5840)$ and $B_{J}(5970)$ were observed by CDF~\cite{Aaltonen:2013atp} and LHCb~\cite{Aaij:2015qla}. The properties of these hadrons have drawn extensive attractions in recent years~\cite{Godfrey:2015dva,Song:2015fha,Chen:2016spr,Gupta:2018zlg,Kher:2017wsq,Gandhi:2019hip,Gupta:2018xds,Godfrey:2019cmi}.

Here, we extend our approach to the effective field theory of excited heavy-light mesons with chiral partner structure~\cite{Nowak:1993vc}. Special interests are given to the states with quantum numbers $j_l^P=\frac{3}{2}^+$  and $j_l^P=\frac{3}{2}^-$ to lay the foundation for
researches on arbitrary spin heavy-light mesons.


The remaining part of this paper is organized as follows. In Sec.~\ref{sec:HeavyEFT}, for convenience, we give the general form of the chiral Lagrangian of the excited heavy-light mesons. In Sec.~\ref{sec:EFTQCD} we derive the excited heavy-light meson Lagrangian from QCD and determine the expression of low energy constants. The numerical results calculated by using the quark self-energy obtained from a typical Dyson-Schwinger equation and from lattice QCD are given in Sec.~\ref{sec:Num}. Section~\ref{sec:dis} is devoted to discussions. The expressions of the LECs with the contribution from the renormalization factor of quark wave function are given in Appendix~\ref{sec:AppA}.

\section{chiral effective Lagrangian for excited heavy-light mesons}

\label{sec:HeavyEFT}

For the convenience of the following description, we present the chiral effective Lagrangian for excited heavy-light meson doublets $T^\mu =(1^+, 2^+)$ and $R^\mu = (1^-, 2^-)$ here. They correspond to the $j_l^P = \frac{3}{2}^+$ in P-wave and $j_l^P = \frac{3}{2}^-$ in D-wave, respectively.  The excited heavy-light meson doublets $T^\mu$ and $R^\mu$ can be expressed as~\cite{Nowak:1993vc}
\begin{eqnarray}
T^\mu(x) & = & \frac{1+\slashed v}{2}\left\{  P^{\ast \mu\nu}_{2^+}\gamma_\nu-\sqrt{\frac{3}{2}}P^{\nu}_{1^+}\gamma_5[g^\mu_\nu-\frac{1}{3}\gamma_\nu(\gamma^{\mu}-v^\mu)] \right\} , \nonumber\\
R^\mu(x) & = & \frac{1+\slashed v}{2}\left\{  P^{\ast\mu\nu}_{2^-}\gamma_\nu\gamma_5-\sqrt{\frac{3}{2}}P_{1^-}^{\nu}[g^\mu_\nu-\frac{1}{3}\gamma_\nu(\gamma^{\mu}+v^\mu)] \right \}, \nonumber\\
\end{eqnarray}
where $(P^\nu_{1^+},P_{2^+}^{\ast\mu\nu})$ refer to $J^P=(1^+, 2^+)$ states, and $(P^\nu_{1^-},P_{2^-}^{\ast\mu\nu})$ refer to $J^P = (1^-, 2^-)$ states, respectively. $v^\mu$ is the velocity of an on-shell heavy quark, i.e. $p^\mu=m_Q v^\mu$ with $v^2=1$.
Then the chiral effective Lagrangian for excited heavy-light mesons can be written as~\cite{Kilian1992hq,Nowak:1993vc}
\begin{eqnarray}
{\cal L} & = & {\cal L}_T + {\cal L}_R + {\cal L}_{TR},
\label{eq:LHHChPT}
\end{eqnarray}
where
\begin{eqnarray}
{\cal L}_T & = & {}-i{\rm Tr} \left(\bar{T}^\mu v\cdot \nabla T_\mu\right) - g_T {\rm Tr} \left(  \bar{T}^\mu T_\mu \gamma^\nu\gamma_5 A_\nu \right) \nonumber\\
& &{} + m_T{\rm Tr} \left(\bar{T}^\mu T_\mu\right),\nonumber\\
{\cal L}_R & = &{} - i {\rm Tr} \left(\bar{R}^\mu v\cdot \nabla R_\mu\right) + g_R {\rm Tr} \left( \bar{R}^\mu R_\mu \gamma^\nu\gamma_5 A_\nu   \right) \nonumber\\
& &{} + m_R{\rm Tr} \left(\bar{R}^\mu R_\mu\right), \nonumber\\
{\cal L}_{TR} & = & g_{TR}{\rm Tr} \left(\bar{R}^\mu T_\mu \gamma^\nu\gamma_5 A_\nu  \right) + h.c..
\end{eqnarray}
The covariant derivative $\nabla_\mu = \partial_\mu - i V_\mu$ with $V_\mu= \frac{i}{2}\left(\Omega^\dagger \partial_\mu \Omega + \Omega \partial_\mu \Omega^\dagger\right)$, and $A_\mu = \frac{i}{2}\left(\Omega^\dagger \partial_\mu \Omega - \Omega \partial_\mu \Omega^\dagger\right)$. The $\Omega$ field is related to the chiral field $U(x) = \exp(i\pi(x)/f_\pi)$ through $U = \Omega^2$. The parameters $g_T, g_R, g_{TR}, m_T$, and $m_R$ are the LECs of the Lagrangian. They are free ones at the level of the effective theory.

As mentioned previously, the states associated with $T^\mu$ and $R^\mu$ are called chiral partners, and their mass splitting arises from the chiral symmetry breaking. In the $D$ meson family, $T^\mu$ may be associated to $T^\mu = (1^+,2^+)=(D_1(2430), D_2^*(2460))$ and a possible identification of the $R^\mu$ doublet may be $R^\mu = (1^-, 2^-)=(D_J^\ast(2600), D(2740))$ with respect that the fact that the states in the $R^\mu$ doublet can decay to their chiral partners in the $T^\mu$ doublet so that should have broad widths , so the spin-averaged masses of the chiral partners are~\cite{ZylaPDG,Aaij:2019sqk}
\begin{eqnarray}\label{eq:EXmass1}
 m_T  \simeq  ~2448~{\rm MeV}, \;\;\;  m_R \simeq 2703~{\rm MeV},
\end{eqnarray}
which yields the mass splitting
\begin{eqnarray}
\Delta m & = & m_R - m_T \simeq 255~{\rm MeV}.
\label{eq:splitting}
\end{eqnarray}

In the $B$ meson family, $T^\mu$ may be associated to $(B_1(5721), B_2^*(5747))$, so the spin-averaged mass is~\cite{ZylaPDG}
\begin{eqnarray}\label{eq:EXmass2}
m_T & \simeq & ~5735~{\rm MeV}
\end{eqnarray}
For the $(1^-,2^-)$ doublet, the situation is subtle since the quantum numbers of the possible candidates in PDG are not well determined. With respect to the mass splitting between chiral partners in the charmed meson sector, the masses of the states in the $R^\mu$ doublet should have masses $\sim 6000~$MeV. This means that it is reasonable to identify the quantum numbers of the state $B_J(5970)$ as $1^-$ or $2^-$ or maybe the $1^-$ and $2^-$ states have nearly degenerate masses.




\section{chiral effective lagrangian of excited heavy-light mesons from QCD}
\label{sec:EFTQCD}

In this section, we follow our previous work~\cite{Chen2020jiq} to derive the chiral effective Lagrangian for excited heavy-light mesons with $j_l^P=\frac{3}{2}^{\pm}$ and its low energy constants.

The generating functional of QCD with an external source $J(x)$ is
\begin{eqnarray}
Z[J] & = & \int \mathcal{D}q \mathcal{D} \bar q \mathcal{D}Q \mathcal{D} \bar Q \mathcal{D} G_{\mu}\varDelta_F(G_{\mu})\nonumber\\
& &\times \exp\left\{i\int d^4 x\left[ {\mathcal{L}_{\mathrm{QCD}}(q,\bar q,Q, \bar Q ,G_{\mu})+\bar q Jq}\right]\right\},\nonumber\\
\label{generating0}
\end{eqnarray}
where $q(x), Q(x)$ and $G_\mu(x)$ are the light-quark, heavy-quark and gluon fields, respectively.

By integrating in the chiral field $U(x)$ and the heavy-light meson fields, and integrating out gluon fields and quark fields, we obtain the effective action for the chiral effective theory as\cite{Chen2020jiq}
\begin{eqnarray}
\label{action2}
S[U,\Pi_2,\bar{\Pi}_2] & = &
{} -i N_c \mathrm{Tr}\ln\left\{\left(i\slashed{\partial}-\bar{\Sigma}\right)I_1 +J_\Omega \right.\nonumber\\
& &\left. {} \qquad\qquad\qquad + \left(i \slashed{\partial}+M\slashed v-M\right)I_4\right.\nonumber\\
& &\left. {} \qquad\qquad\qquad  -\Pi_2-\bar{\Pi}_2 \right\},
\end{eqnarray}
where $\Pi_2$ and $\bar{\Pi}_2$ are the heavy-light meson field and its conjugate, respectively.
$M$ is the mass matrice of heavy quark $Q$. $\bar{\Sigma}$ is the self-energy of the light quark propagator. $I_1=\mathrm{diag}(1,1,0)$ and $I_4=\mathrm{diag}(0,0,1)$ are matrices in the flavor space. $J_\Omega$ is the chiral rotated external source. To obtain this effective action from QCD, we have taken the chiral limit, the heavy quark limit and the large $N_c$ limit. More details can be found in the Appendix A of Ref.~\cite{Chen2020jiq}.

Since the introduced heavy-light meson field $\Pi_2$ is a bilocal field, we need a suitable localization on $\Pi_2$ fields to get a local effective Lagrangian. Here we follow the approach in Ref.~\cite{Nowak:1993vc}, and take the following localization conditions
\begin{eqnarray}
\Pi_2(x_1,x_2) & = & \Pi_2(x_1)\delta(x_1-x_2) +\Pi^\mu_{2} (x_1)\overrightarrow{\partial}_\mu \delta(x_1-x_2) , \nonumber\\
\bar \Pi_2 (x_1,x_2) & = & \bar\Pi_2(x_1)\delta(x_1-x_2)+\delta(x_1-x_2)\overleftarrow{\partial}_\mu  \bar \Pi^\mu_{2}(x_1).\nonumber\\
\end{eqnarray}
where
\begin{eqnarray}
\Pi_{2}(x) & = & H(x)+G(x),\nonumber\\
\Pi^\mu_{2}(x) & = & T^\mu(x)+R^\mu(x),
\end{eqnarray}
Here, the fields $H$ and $G$ refer to $J^P=(0^-, 1^-)$ states and  $J^P = (0^+, 1^+)$ states, the fields $T^\mu$ and $R^\mu$ refer to $J^P=(1^+, 2^+)$ states and $J^P = (1^-, 2^-)$ states respectively. Since we are interested in the heavy-light meson doublets with $j_l^P = \frac{3}{2}^{\pm}$ in this work, we focus on the chiral effective Lagrangian for $\Pi_2^\mu$ alone. So the chiral effective action is reduced to
\begin{eqnarray}
\label{action3}
S[U,\Pi^\mu_2,\bar{\Pi}^\nu_2] & = &
{} -i N_c \mathrm{Tr}\ln\left\{\left(i\slashed{\partial}-\bar{\Sigma}\right)I_1 +J_\Omega \right.\nonumber\\
& &\left. {} \qquad\qquad\qquad + \left(i\slashed{\partial}+M\slashed v-M\right)I_4\right.\nonumber\\
& &\left. {} \qquad\qquad\qquad  -\Pi^\nu_{2}\overrightarrow{\partial}_\nu -  \overleftarrow{\partial}_\mu \bar \Pi^\mu_{2} \right\}
\end{eqnarray}

In order to obtain the chiral effective Lagrangian, we need expand the action $S[U,\Pi^\mu_2,\bar{\Pi}^\nu_2]$ with respect to the fields $U$, $\Pi^\mu_2$  and $\bar{\Pi}^\nu_2$. The kinetic terms of the excited heavy-light meson fields $T^\mu$ (a similar equation holds for $R^\mu$) arise from
\begin{eqnarray}
S_{2} & \equiv & \frac{\delta^2 S}{\delta \bar{T}^\mu \delta T^\nu} T^\nu\bar{T}^\mu\nonumber\\
& = & iN_c\int d^4x_1 d^4 x_2 \nonumber\\
& & {} \times \mathrm{Tr}\Bigg[\left(i\slashed{\partial} -\bar{\Sigma}\right)^{-1}\delta(x_2-x_1)\overleftarrow{\partial}_\mu\bar{T}^\mu(x_1)\nonumber\\
&&\qquad\qquad \times\left(i v\cdot\partial \right)^{-1} \overrightarrow{\partial}_\nu T^\nu(x_2)\delta(x_1-x_2)\Bigg],\nonumber\\
\label{S2}
\end{eqnarray}
Taking the derivative expansion up to the first order, we obtain
\begin{widetext}
\begin{eqnarray}
S_{2} & = & i N_c\int d^4x \int\frac{d^4  p}{(2\pi)^4} \left[-\frac{p^2-(v\cdot p)^2}{ p^2-\Sigma^2} +\frac{p^2\Sigma}{( p^2-\Sigma^2)v\cdot  p}\right] \mathrm{Tr}\left[\bar{T}^\mu(x)  T_\mu(x) \right]\notag\\
& & {} ~~~~+iN_c\int d^4x \int\frac{d^4  p}{(2\pi)^4} \left[-\frac{p^2}{( p^2-\Sigma^2) v\cdot  p}- \frac{[2\Sigma +2\Sigma'(p^2+\Sigma^2)][p^2-(v\cdot p)^2]} {( p^2-\Sigma^2)^2}\right]\mathrm{Tr}\left[ T_\mu(x)   v \cdot \partial \bar{T}^\mu (x)  \right]\notag\\
& &{} ~~~~+iN_c\int d^4x \int\frac{d^4  p}{(2\pi)^4} \left[-\frac{2\Sigma'( p^2+\Sigma^2)[p^2-(v\cdot p)^2]}{( p^2-\Sigma^2)^2}\right]\mathrm{Tr}\left[ T_\mu (x)v\cdot V_\Omega \bar T ^\mu (x)\right].
\label{eq:S2}
\end{eqnarray}
In the calculation, we have used the relation $T^\mu \slashed v =-T^\mu$. It is easy to identify the mass term and the kinetic term in Eq.~\eqref{eq:S2}. In addition, an interaction term  between the $T^\mu$ fields and $ V_\Omega $ fields also appear in Eq.~\eqref{eq:S2} because we have taken $\bar{\Sigma}(x-y)=\Sigma(\nabla^2)\delta(x-y)$ in the calculation to retain the correct chiral transformation properties in the theory~\cite{Yang:2002hea}.

The interaction between the heavy-light meson field $T^\mu$ and the light Goldstone boson field $S_{3}$ can also be obtained by expanding the action $S[U,\Pi^\mu_2,\bar{\Pi}^\nu_2]$ with respect to the external $J_\Omega$, $\Pi^\mu_2$  and $\bar{\Pi}^\nu_2$ as
\begin{eqnarray}\label{S3}
S_{3} & \equiv & \frac{\delta^3 S}{\delta J_\Omega  \delta \bar T^\mu \delta  T^\nu } J_\Omega T^\nu\bar{T}^\mu\notag\\
& = &{} -iN_c\int d^4x_1d^4x_2d^4x_3\nonumber\\
&&\qquad {} \times \mathrm{Tr}\left[(i\slashed{\partial} -\Sigma)^{-1}\delta (x_2-x_3) J_{\Omega}(x_3) (i\slashed{\partial} -\Sigma)^{-1} \delta(x_3-x_1) \overleftarrow{\partial}_\mu\bar T^\mu(x_1)\left(i v\cdot\partial\right)^{-1} \delta(x_1-x_2) \overrightarrow{\partial}_\nu  T^\nu (x_2)\right].
\label{S3}
\end{eqnarray}
Upto the first order of the derivative expansion, one obtains
\begin{eqnarray}
S_{3} & = &{} -iN_c\int d^4x \int\frac{d^4 p}{(2\pi)^4}\left[\frac{p^2(\Sigma^2+\frac{2}{3}p^2)}{(p^2-\Sigma^2) v\cdot p} - \frac{2\Sigma[p^2-(v\cdot p)^2]}{(p^2-\Sigma^2)^2 } \right] \mathrm{Tr}\left[  \bar{T}(x) ^\mu T_\mu(x)  \gamma^\nu\gamma_5 A_\nu\right]\notag\\
& &{} -iN_c\int d^4x \int\frac{d^4 p}{(2\pi)^4} \left[\frac{p^2}{(p^2-\Sigma^2) v\cdot p} +\frac{2\Sigma [p^2-(v\cdot p)^2] }{(p^2-\Sigma^2)^2} \right]  \mathrm{Tr}\left[  T_\mu(x) v \cdot V_{\Omega}\bar T^\mu(x)  \right].
\label{eq:SJHH}
\end{eqnarray}
\end{widetext}

Summing up Eqs.~\eqref{eq:S2} and \eqref{eq:SJHH}, we obtain the expressions of the constants $m_T$ and $g_T$ as
\begin{eqnarray}
m_T & = & \frac{iN_c}{Z_T}  \int\frac{d^4 p}{(2\pi)^4}\left[\frac{p^2-(v\cdot p)^2}{ p^2-\Sigma^2} -\frac{p^2\Sigma}{( p^2-\Sigma^2)v\cdot  p}\right],  \nonumber\\
g_{T} & = &{} -\frac{iN_c}{Z_T}\int\frac{d^4 p}{(2\pi)^4}\left[\frac{p^2(\Sigma^2+\frac{2}{3}p^2)}{(p^2-\Sigma^2) v\cdot p} - \frac{2\Sigma[p^2-(v\cdot p)^2]}{(p^2-\Sigma^2)^2 } \right] ,
\nonumber\\
\label{eq:mHgH}
\end{eqnarray}
with $Z_T$ being the wave function renormalization factor
\begin{eqnarray}
Z_T & = & iN_c\int\frac{d^4 p}{(2\pi)^4} \biggr[{}-\frac{p^2}{( p^2-\Sigma^2) v\cdot  p} \nonumber\\
& &{}\qquad\qquad\qquad - \frac{[2\Sigma +2\Sigma'(p^2+\Sigma^2)][p^2-(v\cdot p)^2]} {( p^2-\Sigma^2)^2}\biggr].\nonumber
\end{eqnarray}

Following the same procedure, one can get the LECs associated to $R^\mu$.
The only difference in the calculation is the relation $R^\mu\slashed v =R^\mu$, which induces the differences between the LECs of $R^\mu$ and those of $T^\mu$. The LECs associated to $T^\mu$ read
\begin{eqnarray}
m_R & = & \frac{iN_c}{Z_R}\int\frac{d^4 p}{(2\pi)^4} \left[\frac{p^2-(v\cdot p)^2}{ p^2-\Sigma^2} +\frac{p^2\Sigma}{( p^2-\Sigma^2)v\cdot  p}\right], \nonumber\\
g_{R} & = &{} -\frac{iN_c}{Z_R}\int\frac{d^4 p}{(2\pi)^4}\left[\frac{p^2(\Sigma^2+\frac{2}{3}p^2)}{(p^2-\Sigma^2) v\cdot p} - \frac{2\Sigma[p^2-(v\cdot p)^2]}{(p^2-\Sigma^2)^2 } \right],
\nonumber\\
Z_R & = & iN_c\int\frac{d^4 p}{(2\pi)^4}  \biggr[{}-\frac{p^2}{( p^2-\Sigma^2) v\cdot  p} \nonumber\\
& &{}\qquad\qquad\qquad + \frac{[2\Sigma +2\Sigma'(p^2+\Sigma^2)][p^2-(v\cdot p)^2]} {( p^2-\Sigma^2)^2}\biggr].\nonumber\\
\label{eq:mGgG}
\end{eqnarray}
We also obtain the coupling constant between the chiral partner fields $T^\mu$ and $R^\mu$ as
\begin{eqnarray}
g_{T R} & = & {} - i \frac{N_c}{\sqrt{Z_T Z_R}}\int\frac{d^4 p}{(2\pi)^4}\biggl[\frac{p^2(\Sigma^2+p^2)}{(p^2-\Sigma^2)^2 v\cdot p} \biggr].
\label{eq:gHG}
\end{eqnarray}
In the above expressions, $\Sigma(-p^2)$ stands for self-energy of light quarks to be calculated from QCD. This is an apparent indicator that we have established a connection between the low energy constants of the chiral effective theory and the Green functions of QCD.
\section{Numerical results and discussions}
\label{sec:Num}

In order to obtained the explicit numerical values of the LECs, we first need to obtain the light quark self-energy $\Sigma(-p^2)$. As in our previous works~\cite{Chen2020jiq,Chen2020epl}, we use two method to determine $\Sigma(-p^2)$ for comparison, namely, the gap equation, i.e., the Dyson-Schwinger equation for quarks, and the fittings from lattice QCD.

For the Dyson-Schwinger equation method, we use the differential form of the gap equation~\cite{Yang:2002hea}:
\begin{eqnarray}\label{gap}
& &\left(\frac{\alpha_s(x)}{x}\right)'\Sigma(x)''-\left(\frac{\alpha_s(x)}{x}\right)'' \Sigma(x)' \nonumber\\
&&\qquad\qquad\qquad\quad {}- \frac{3C_2(R)}{4\pi}\frac{x\Sigma(x)}{x+\Sigma^2(x)}
\left(\frac{\alpha_s(x)}{x}\right)'^2=0,\nonumber
\end{eqnarray}
where the boundary conditions are
\begin{eqnarray}
\Sigma'(0) & = &{} - \frac{3C_2(R)\alpha_s(0)}{8\pi\Sigma(0)}, \nonumber\\
\Sigma(\Lambda') & = & \frac{3C_2(R)\alpha_s(\Lambda')}{4\pi\Lambda'}
\int^{\Lambda'}_0dx\frac{x\Sigma(x)}{x+\Sigma^2(x)},
\end{eqnarray}
with $\alpha_s$ being the running coupling constant of QCD. $\Lambda'$ is an ultraviolet cutoff regularizing the integral, which is taken $\Lambda^\prime \rightarrow \infty$ eventually. To solve the Eq.~\eqref{gap}, we take a model description for $\alpha_s$ given in Ref.~\cite{Dudal:2012zx}
\begin{equation}
\alpha_s(p^2)=\alpha_0p^2\frac{p^2+M^2}{p^4+(M^2+m^2)p^2+\lambda^4}.
\end{equation}
For convenience, we call it Gribov-Zwanziger (G-Z) Model. The parameters are $M^2=4.303~{\rm GeV}^2 , (M^2+m^2)= 0.526~{\rm GeV}^2 $ and $\lambda^4=0.4929~{\rm GeV}^4$~\cite{Dudal:2012zx}. $\alpha_0$ is a model parameter to be determined. Although the UV behavior of G-Z formalism  is inconsistent with QCD, it should not have a sizable impact on our results because the LECs are mostly controlled by its low energy behavior.

\begin{figure}[h]
	\centering	
	\includegraphics[height=5.0cm]{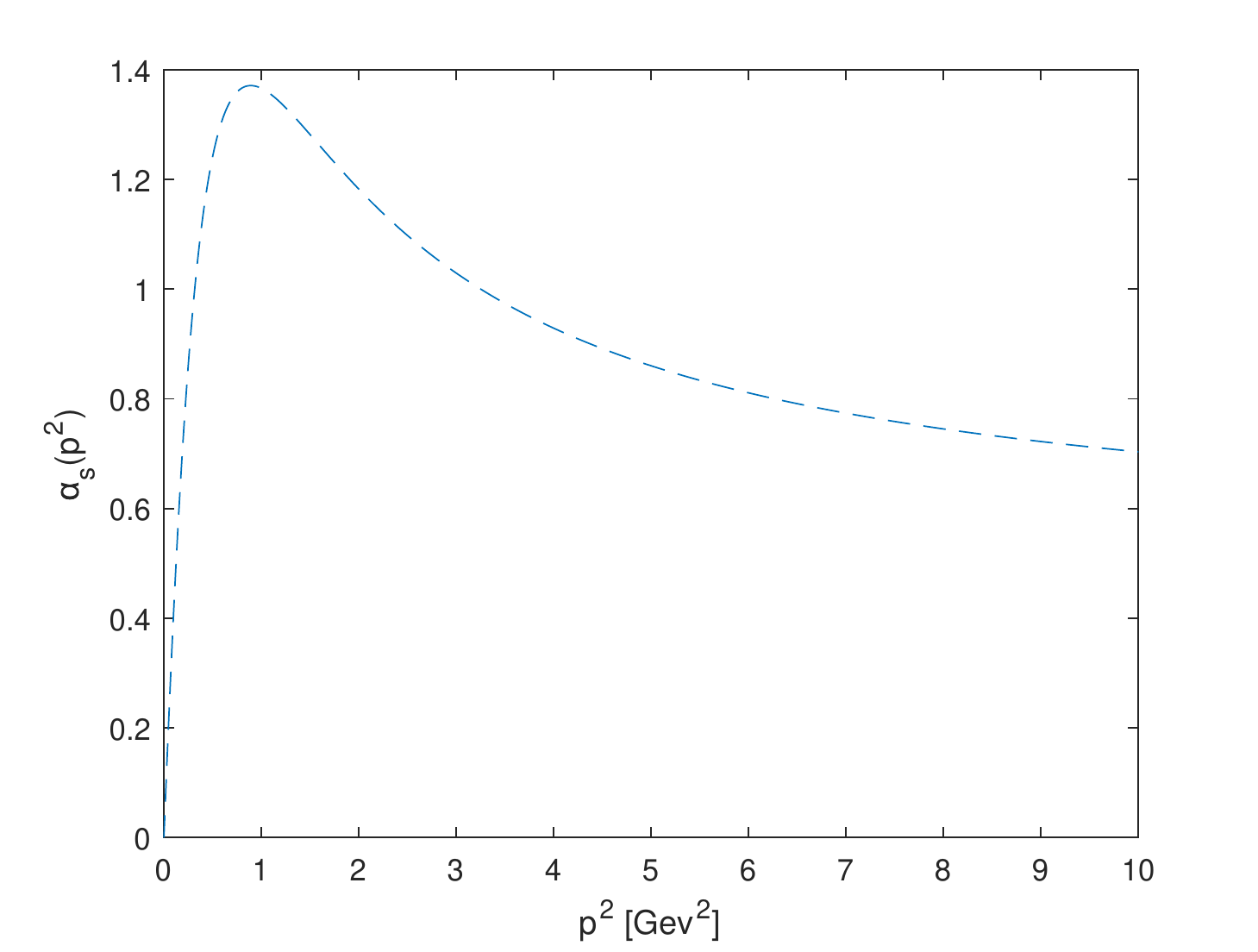}
	\caption{Running coupling constant of the G-Z model with $a_0=0.52$. }	
	\label{rcc}
\end{figure}

\begin{figure}[h]
	\centering	
	\includegraphics[height=5cm]{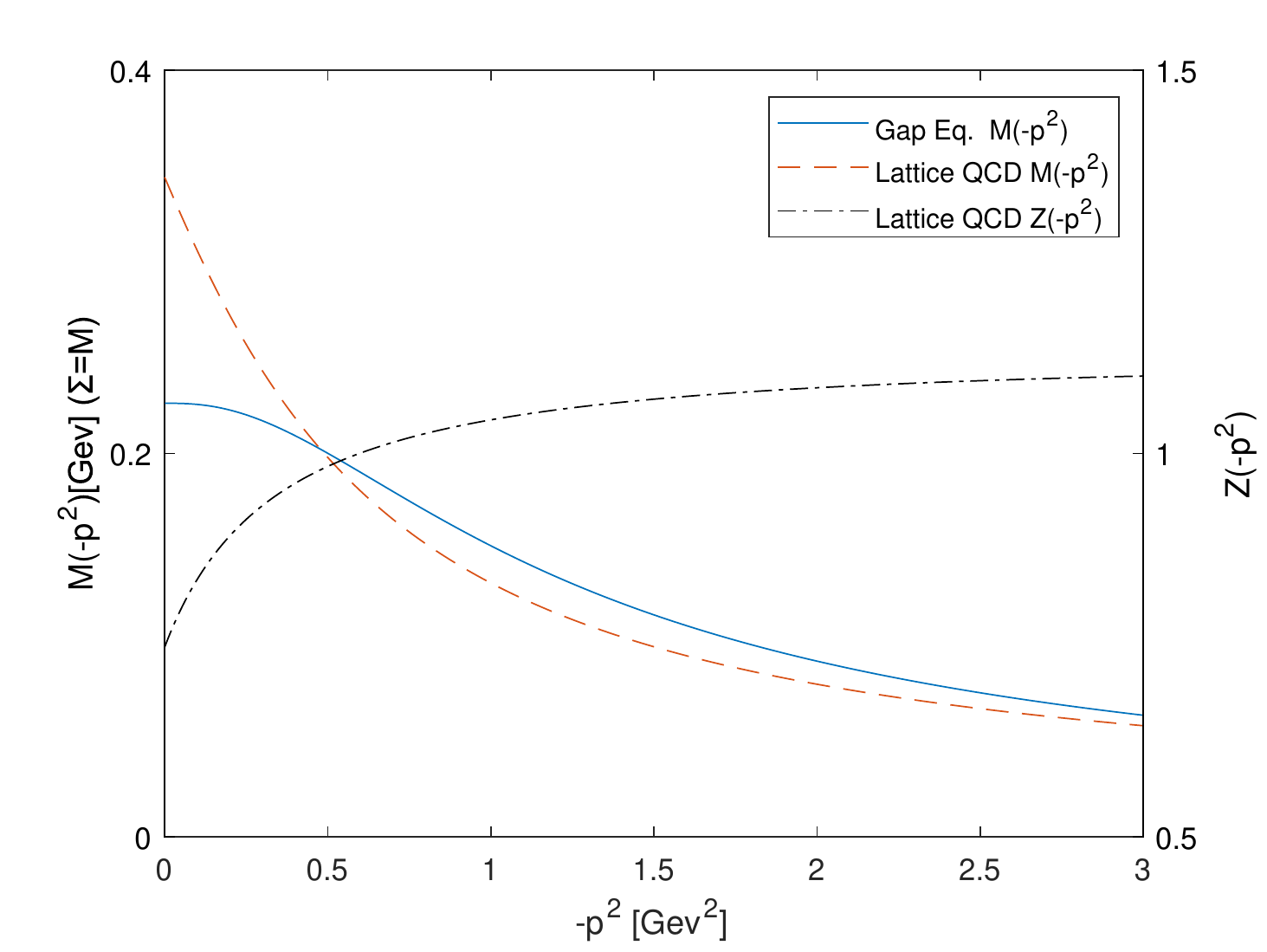}
	\caption{The lattice fittings of $M(-p^2)$ and $Z(-p^2)$ given in Ref. \cite{Oliveira:2018ukh} and $\Sigma(-p^2)$ from the gap equation with G-Z Model. }	
	\label{selfenergy}
\end{figure}

Solving the gap equation, we obtain $\Sigma(-p^2)$. Then the LECs are calculated according to Eqs.~\eqref{eq:mHgH}-\eqref{eq:gHG}. It is clear that the integrals of the LECs have a physical ultraviolet cutoff $\Lambda_c$ which should be of the order of the chiral symmetry breaking scale and serves as another parameter in our calculations. Since we are studying excited states, the cutoff $\Lambda_c$ should take a bit larger value than 1 GeV. For the excited states we are considering now, the energy of the light quark in the chromoelectrical fields generated by the heavy quark could be up to $\sim1.45$ GeV, so $\Lambda_c$ taking a value $\gtrsim 1.5$ GeV would be more appropriate for the present situation. For the G-Z model, we take $\Lambda_c = 1.6$~GeV and determine the model parameter $a_0$ by requiring the calculated $f_\pi$ be consistent with the corresponding experimental value. We find that using $a_0=0.52$ we can obtain the well-established quantity $f_\pi$. The results are listed in Table~\ref{tab1}. For comparison, we also list the numerical results obtained with different $a_0$'s in the table. To give an intuitive impression, we draw the running coupling constant $\alpha_s$ calculated with the G-Z formalism at $a_0=0.52$ in Fig.~\ref{rcc}. The light quark self-energy $\Sigma(-p^2)$ solved by the gap equation \eqref{gap} is shown in Fig.~\ref{selfenergy}.
\begin{table*}[!tbp]
	\centering
	\caption{The heavy-light meson masses and the coupling constants calculated from the G-Z Model with $\Lambda_c=1.6$ GeV. }
		\setlength{\tabcolsep}{3mm}
	\begin{tabular}{ccccccccccc}
		\hline\hline
			$a_0$  & $\Sigma(0)$~(GeV) & $f_{\pi}$(GeV) &  ${}-\langle\bar{\psi}\psi\rangle$ (GeV$^{3}$) & $g_R $  & $g_T$ & $g_{TR}$ & $m_R$~(GeV) & $m_T$~(GeV) & $\Delta m $~(GeV) \\
			\hline
    	0.50   & 0.158 & 0.073 & $(0.328)^3$ & 0.882 & -0.479 & -0.990 & 1.365 & 1.104 & 0.261 \\
    	0.51  & 0.193 & 0.084 & $(0.350)^3$ & 0.936 & -0.442 & -0.987 & 1.411 & 1.091 & 0.321 \\
    	0.52  & 0.226 & 0.093 & $(0.368)^3$ & 0.991 & -0.409 & -0.984 & 1.459 & 1.080 & 0.379 \\
    	0.53  & 0.259 & 0.102 & $(0.385)^3$ & 1.047 & -0.377 & -0.981 & 1.510 & 1.071 & 0.439 \\
    	0.54  & 0.291 & 0.109 & $(0.399)^3$ & 1.107 & -0.347 & -0.979 & 1.564 & 1.064 & 0.500 \\
		
			\hline	\hline
	\end{tabular}%
	\label{tab1}%
\end{table*}%

Since there are more and more excited states observed experimentally whose masses and quantum numbers have not been confirmed yet, we focus on the masses of the chiral partners in our method in this work and hopefully could help identifying some excited mesons.
The mass splitting $\Delta m$ is a direct indicator of the chiral symmetry breaking. In our calculations, it is completely originated from the non-vanishing of the quark running mass, which can be reflected by the quark condensate or the value of $\Sigma(0)$. From Table~\ref{tab1}, one can see that $\Delta m$ increases as the quark condensate or $\Sigma(0)$ increases. At $a_0=0.52$, we find $\Delta m=379$ MeV which is a bit larger than the expected value $\sim 255$ MeV as given in ~\eqref{eq:EXmass1}. This is not surprising because on one hand the measured masses of the relevant excited mesons are still not accurate, and on the other hand our results are suffering from uncertainties due to the ignorance of the $1/m_Q$ contributions, etc..

The masses $m_T$ and $m_R$ listed in Table~\ref{tab1} are the ``residue masses" with the heavy quark mass rotated away. The physical masses (denoted as $\tilde{m}_T$ and $\tilde{m}_R$) can be obtained by simply adding back the heavy quark mass to $m_T$ and $m_R$. For the $D$ meson sector, using $m_c\approx 1.27$ GeV~\cite{ZylaPDG}, we obtain the physical masses to be $\tilde{m}_T\approx 2.35$ GeV and $\tilde{m}_R\approx 2.73$ GeV with $a_0=0.52$. Keeping in mind the large uncertainties from both the theoretical side and the experimental side, these values are roughly comparable to the corresponding experimental data~\eqref{eq:EXmass1}.
For the $B$ meson family,  using $m_b\approx 4.66$ GeV ~\cite{ZylaPDG},  we obtain the physical masses to be $\tilde{m}_T\approx 5.74$ GeV and $\tilde{m}_R\approx 6.12$ GeV.  Comparing with the experimental data for the bottom mesons in~\eqref{eq:EXmass2}, the numerical result of $\tilde{m}_T$ is close to the corresponding experimental data. And we expect $(1^-,2^-)$ $B$ meson states appear at around 6.12 GeV. Actually, the excited state $B_J(5970)$ reported in Refs.~\cite{Aaltonen:2013atp,Aaij:2015qla} with the mass $\sim5970$ MeV could be the candidate for the $1^-$ or $2^-$ states. Up to now, experiments have not yet determined the $J^P$ quantum numbers for $B_J(5970)$. We expect this could be done in the near future.

Now, let us turn to the results using the quark self-energy fitted from lattice QCD. We use the lattice fittings for the quark wave function renormalization $Z(-p^2)$ and the running quark mass $M(-p^2)$ given in Ref.~\cite{Oliveira:2018ukh}. The formula of LECs shown in Eqs.~\eqref{eq:mHgH}-\eqref{eq:gHG} can be directly extended to incorporate the contributions from $Z(-p^2)$. The relevant expressions are given in Appendix A. The functions $Z(-p^2)$ and $M(-p^2)$ are also plotted in Fig.~\ref{selfenergy}.
The numerical results of LECs calculated by using these lattice fittings are given in Table~\ref{tab2}. There is no free parameters in the functions of $M(-p^2)$ and $Z(-p^2)$. We display results from various values of the cutoff $\Lambda_c$ for comparison.

\begin{table*}[!htbp]
	\centering
	\caption{LECs calculated from lattice fittings given in Ref \cite{Oliveira:2018ukh}. }
	\begin{tabular}{cccccccccc}
		\hline	\hline
		$\Lambda_c$(GeV)  & $M(0)$(GeV) 	& $f_{\pi}$ (GeV) & ${}-\langle\bar{\psi}\psi\rangle$ (GeV$^{3}$) & $\quad$ $g_R$ $\quad$  & $\quad$ $g_T$ $\quad$ & $\quad$ $g_{TR}$ $\quad$ & $m_R$~(GeV) & $m_T$~(GeV) & $\Delta m $~(GeV)\\
		\hline
	    1.4   & 0.344 & 0.095 & $(0.341)^3$ & 1.101 & $-0.398$ & $-1.031$ & 1.349 & 0.948 & 0.401 \\
    	1.5   & 0.344 & 0.096 & $(0.351)^3$ & 1.054 & -0.432 & -1.041 & 1.405 & 1.021 & 0.384 \\
    	1.6   & 0.344 & 0.096 & $(0.359)^3$ & 1.015 & -0.462 & -1.050 & 1.462 & 1.095 & 0.368 \\
    	1.7   & 0.344 & 0.096 & $(0.367)^3$ & 0.983 & -0.488 & -1.058 & 1.522 & 1.169 & 0.352 \\
    	1.8   & 0.344 & 0.096 & $(0.375)^3$ & 0.956 & -0.511 & -1.064 & 1.582 & 1.245 & 0.337 \\
		\hline	\hline
	\end{tabular}%
	\label{tab2}%
\end{table*}%

From Table~\ref{tab2}, one can see that the numerical results of LECs are comparable to the results from the G-Z model when $\Lambda_c=1.6$~GeV, which implies that $Z(-p^2)$ does not have significant impacts on the excited states. The physical masses are $\tilde{m}_T\approx 2.37$ GeV and $\tilde{m}_R\approx 2.73$ GeV for excited $D$ mesons, and $\tilde{m}_T\approx 5.76$ GeV and $\tilde{m}_G\approx 6.12$ GeV for excited $B$ mesons. The conclusions could be drawn from these results are the same as we have in the G-Z model.

\section{Conclusions}
\label{sec:dis}
In this paper, we derive the chiral effective Lagrangian for excited heavy-light mesons from QCD. We focus on the chiral partners with $j_l^P=\frac{3}{2}^+$ and $j_l^P=\frac{3}{2}^-$ which amounts to ($1^+,2^+$) and ($1^-,2^-$) states respectively. The low energy constants are expressed as momentum integrals of the light quark self-energy, which in turn are obtained by using the gap equation or lattice QCD. The relevant LECs in the chiral Lagrangian are calculated, and the resulted masses for the excited $D$ mesons are found roughly consistent with experimental data. For the excited $B$ mesons, we obtain $\tilde{m}_T\approx 5.74$ GeV and $\tilde{m}_R\approx 6.12$ GeV. We find that $\tilde{m}_T$ are consistent to the masses of $(B_1(5721), B_2^*(5747))$, and the result of $\tilde{m}_R$ suggests that $B_J(5970)$ could be a good candidate for the states in the doublet $(1^-,2^-)$.

As more and more excited states of charmed mesons and bottom mesons are discovered in experiments, it will be an interesting and important research to extend our method to obtain the chiral effective Lagrangian of heavy-light mesons with arbitrary spins. Since this extension is straightforward, we will not go to details here.

We finally want to say that, so far, we did not discuss the chiral partner structure of the heavy-light mesons including a strange quark. One of the reasons is that the quark content of some of these mesons is still under debate. For example, the mesons $D_{s0}(2317)$ and $D_{s1}(2460)$ which can be arranged to the $G$ doublet with quantum numbers $G = (0^+,1^+)$ are regarded as the chiral partners of $D$ and $D^\ast$, respectively, in the $H$ doublet with quantum numbers $H=(0^-,1^-)$~\cite{Nowak:2003ra}. The molecular state interpretation of these mesons and their bottom cousins cannot be ruled out~\cite{Faessler:2007gv,Faessler:2007us,Faessler:2008vc}. We leave the discussion of these strange mesons as future works.


\acknowledgments
The work of Y.~L. M. was supported in part by National Science Foundation of China (NSFC) under Grant No. 11875147 and No.11475071. H.-F. Fu was supported by NSFC under Grant No.12047569. Q. W. was supported by the National Science Foundation of China (NSFC) under Grant No. 11475092.

\appendix

\renewcommand{\appendixname}{Appendix}

\section{FORMULA FOR LECS WITH $Z(-p^2)$}

\label{sec:AppA}

The quark propagator can be generally written as
\begin{eqnarray}
S(p) & = &\frac{i}{A(-p^2) \slashed  p\ -  B(-p^2) } \nonumber\\
& = & i \, Z(-p^2) \, \frac{ \slashed  p\ + M(-p^2) }{ p^2 - M^2(-p^2)},\nonumber
\end{eqnarray}
where $Z(-p^2) = 1/A(-p^2)$ stands for the quark wave function renormalization and $M(-p^2) = B(-p^2) / A(-p^2)$ is the renormalization group invariant running quark mass. After a series of calculations, we can get the LECs with $Z(-p^2)$ as follows:
\begin{widetext}
\begin{eqnarray}
m_T & = & \frac{iN_c}{Z_T}  \int\frac{d^4 p}{(2\pi)^4}  \biggr[{}\frac{p^2-(v\cdot p)^2}{p^2-M^2} - \frac{p^2 M}{(p^2-M^2)v\cdot p}\biggr]Z,  \nonumber\\
g_{T} & = & -\frac{iN_c}{Z_T}\int\frac{d^4 p}{(2\pi)^4}\biggl[\frac{p^2(M^2+\frac{2}{3}p^2)}{(p^2-M^2)^2 v\cdot p}  - \frac{2M[p^2-(v\cdot p)^2]}{(p^2-M^2)^2 } \biggr]Z^2,\nonumber\\
m_R & = & \frac{iN_c}{Z_R}\int\frac{d^4 p}{(2\pi)^4} \biggr[{}\frac{p^2-(v\cdot p)^2}{p^2-M^2} + \frac{p^2 M}{(p^2-M^2)v\cdot p}\biggr]Z, \nonumber\\
g_{R} & = & -\frac{iN_c}{Z_R}\int\frac{d^4 p}{(2\pi)^4}\biggl[\frac{p^2(M^2+\frac{2}{3}p^2)}{(p^2-M^2)^2 v\cdot p}  + \frac{2M[p^2-(v\cdot p)^2]}{(p^2-M^2)^2 } \biggr]Z^2,\nonumber\\
Z_T & = & iN_c\int\frac{d^4 p}{(2\pi)^4} \Biggr\{ \biggr[-\frac{p^2}{(p^2-M^2) v\cdot p} - \frac{[2M+2Z M'(p^2+M^2)][p^2-(v\cdot p)^2]}{(p^2-M^2)^2}  \Biggr ] Z -\frac{2Z' M }{p^2-M^2} \biggr\}\nonumber\\
Z_R & = &iN_c\int\frac{d^4 p}{(2\pi)^4} \Biggr\{ \biggr[-\frac{p^2}{(p^2-M^2) v\cdot p} +\frac{[2M+2Z M'(p^2+M^2)][p^2-(v\cdot p)^2]}{(p^2-M^2)^2}  \Biggr ] Z +\frac{2Z' M }{p^2-M^2} \biggr\},\nonumber\\
g_{T R} & = &  - i \frac{N_c}{\sqrt{Z_T Z_R}}\int\frac{d^4 p}{(2\pi)^4}\biggl[\frac{p^2(M^2+p^2)}{(p^2-M^2)^2 v\cdot p} \biggr]Z^2.
\end{eqnarray}
\end{widetext}
%

\bibliography{paper}

\end{document}